\documentclass{nature}

\usepackage{amsmath}
\usepackage{amssymb}
\usepackage{graphicx}
\usepackage{upgreek}
\usepackage{float}

\title{A Fast Radio Burst Host Galaxy}

\author{
E. F. Keane$^{1,2,3}$,
S. Johnston$^{4}$,
S. Bhandari$^{2,3}$,
E. Barr$^{2}$,
N. D. R. Bhat$^{5,3}$,
M. Burgay$^{6}$,      
M. Caleb$^{7,2,3}$,   
C. Flynn$^{2,3}$,     
A. Jameson$^{2,3}$,   
M. Kramer$^{8,10}$,   
E. Petroff$^{2,3,4}$, 
A. Possenti$^{6}$,
W. van Straten$^{2}$, 
M. Bailes$^{2,3}$,
S. Burke-Spolaor$^{9}$,
R. P. Eatough$^{8}$,  
B. Stappers$^{10}$,   
T. Totani$^{11}$,     
M. Honma$^{13,14}$,   
H. Furusawa$^{13}$,   
T. Hattori$^{12}$,    
T. Morokuma$^{15,16}$,
Y. Niino$^{13}$,      
H. Sugai$^{16}$,      
T. Terai$^{12}$,      
N. Tominaga$^{17,16}$,
S. Yamasaki$^{11}$,   
N. Yasuda$^{16}$,     
R. Allen$^{2}$,       
J. Cooke$^{2,3}$,     
J. Jencson,$^{18}$    
M. M. Kasliwal$^{18}$,
D. L. Kaplan$^{19}$,  
S. J. Tingay$^{5,3}$,    
A. Williams$^{5}$,    
R. Wayth$^{5,3}$,     
P. Chandra$^{20}$,    
D. Perrodin$^{6}$,    
M. Berezina$^{8}$,    
M. Mickaliger$^{10}$  
\& C. Bassa$^{21}$
}

\begin{document}

\maketitle

\begin{affiliations}
 \item SKA Organisation, Jodrell Bank Observatory, SK11 9DL, UK. 
 \item Centre for Astrophysics and Supercomputing, Swinburne University of Technology, Mail H29, PO Box 218, VIC 3122, Australia. 
 \item ARC Centre of Excellence for All-sky Astrophysics (CAASTRO). 
 \item CSIRO Astronomy \& Space Science, Australia Telescope National Facility, P.O. Box 76, Epping, NSW 1710, Australia 
 \item International Centre for Radio Astronomy Research, Curtin University, Bentley, WA 6102, Australia 
 \item INAF --- Osservatorio Astronomico di Cagliari, Via della Scienza 5, I-09047 Selargius (CA), Italy 
 \item Research School of Astronomy and Astrophysics, Australian National University, ACT 2611, Australia 
 \item Max-Planck-Institut f\"ur Radioastronomie, Auf dem H\"ugel 69, D-53121 Bonn, Germany 
 \item National Radio Astronomy Observatory, Socorro, NM, USA 
 \item Jodrell Bank Centre for Astrophysics, School of Physics and Astronomy, The University of Manchester, Manchester M13 9PL, UK 
 \item Department of Astronomy, the University of Tokyo, Hongo, Tokyo 113-0033, Japan 
 \item Subaru Telescope, National Astronomical Observatory of Japan, 650 North A`ohoku Pl., Hilo, HI 96720, USA 
 \item National Astronomical Observatory of Japan, 2 Chome-21-1 Osawa, Mitaka, Tokyo 181-8588, Japan 
 \item Department of Astronomical Science, SOKENDAI (Graduate University for the Advanced Study), Osawa, Mitaka 181-8588,  Japan 
 \item Institute of Astronomy, Graduate School of Science, The University of Tokyo, 2-21-1 Osawa, Mitaka, Tokyo 181-0015, Japan 
 \item Kavli Institute for the Physics and Mathematics of the Universe (WPI), The University of Tokyo Institutes for Advanced Study, The University of Tokyo, Kashiwa, Chiba 277-8583, Japan 
 \item Department of Physics, Faculty of Science and Engineering, Konan University, 8-9-1 Okamoto, Kobe, Hyogo 658-8501, Japan 
 \item Cahill Center for Astrophysics, California Institute of Technology, 1200 E California Blvd, Pasadena, CA 91125, USA 
 \item Department of Physics, University of Wisconsin-Milwaukee, Milwaukee, WI 53201, USA 
 \item National Centre for Radio Astrophysics, Tata Institute of Fundamental Research, Pune University Campus, Ganeshkhind, Pune 411 007, India 
 \item ASTRON, the Netherlands Institute for Radio Astronomy, Postbus 2, NL-7990 AA Dwingeloo, the Netherlands 
\end{affiliations}


\newpage
\begin{abstract}
In recent years, millisecond duration radio signals originating from distant galaxies appear to have been discovered in the so-called Fast Radio Bursts\cite{lbm+07,kskl12,tsb+13,sch+14,bb14,rsj15,pbb+15,mls+15,cha+16}. These signals are dispersed according to a precise physical law and this dispersion is a key observable quantity which, in tandem with a redshift measurement, can be used for fundamental physical investigations\cite{mcq14,zlw+14}. While every fast radio burst has a dispersion measurement, none before now have had a redshift measurement, due to the difficulty in pinpointing their celestial coordinates. Here we present the discovery of a fast radio burst and the identification of a fading radio transient lasting $\sim 6$ days after the event, which we use to identify the host galaxy; we measure the galaxy's redshift to be $z=0.492\pm0.008$. The dispersion measure and redshift, in combination, provide a direct measurement of the cosmic density of ionised baryons in the intergalactic medium of $\Omega_{\mathrm{IGM}}=4.9 \pm 1.3\%$, in agreement with the expectation from WMAP\cite{wmap13}, and including all of the so-called ``missing baryons''. The $\sim6$-day transient is largely consistent with a short gamma-ray burst radio afterglow\cite{cf12}, and its existence and timescale do not support progenitor models such as giant pulses from pulsars, and supernovae. This contrasts with the interpretation of another recently discovered fast radio burst\cite{mls+15}, suggesting there are at least two classes of bursts.
\end{abstract}

As part of the SUrvey for Pulsars and Extragalactic Radio Bursts (SUPERB) project at the Parkes radio telescope, we perform real-time searches of the sky at high time resolution (see Methods). On 2015 April 18 UTC we detected a fast radio burst (FRB) which we refer to as FRB~150418. Its dispersion measure (DM) is $776.2(5)\;\mathrm{cm}^{-3}\;\mathrm{pc}$, which is 4.1 times the maximum Galactic contribution expected from this line of sight through the Milky Way\cite{cl02}. The UTC at which the FRB was detected is 04:29:07.056 at $1382$~MHz, or 04:29:05.370 at infinite frequency with the dispersion delay removed (Figure 1). The observed pulse width is $0.8\pm0.3$~ms, consistent with dispersion smearing due to the finite frequency resolution of the spectrometer, indicating that the intrinsic width is unresolved. No scattering is evident, consistent with the expected contribution from the Galaxy which is $\ll 1$~ms at this latitude\cite{bcc+04}. The linear polarisation is not large at $8.5\pm1.5\%$ and the circular polarisation is consistent with zero. Given the low level of linear polarisation the rotation measure (RM) is not known precisely and is $\mathrm{RM}=36\pm52\;\mathrm{rad}\;\mathrm{m}^{-2}$. In 13 hours of followup observations no repeat burst was detected (see Methods).

Upon detection of FRB~150418 at Parkes, a network of telescopes was triggered across a wide range of wavelengths (see Methods). Beginning two hours after the FRB, observations with the Australia Telescope Compact Array (ATCA) were carried out at 5.5 and 7.5 GHz, identifying two variable compact sources. One of the variable sources is consistent with a positive spectral index GHz-peaked spectrum source as previously identified in observations at these frequencies\cite{bhh+15}. The other variable source (RA 07:16:34.6, Dec $-$19:00:40), offset by $1.944$~arcmin from the centre of the Parkes beam, was seen at 5.5 GHz at a brightness of $0.27(5)$~mJy/beam just 2 hours after the FRB. The source was then seen to fade over subsequent epochs, settling at a brightness of $\sim0.09(2)$~mJy/beam (Figure 2). The source is also seen at 7.5 GHz at $0.18(3)$~mJy/beam in the first epoch but subsequently not detected. These observations indicate a $\sim 6$-day transient with a negative spectral index; we obtain $\alpha=-1.37$ in the first epoch, for a power-law spectrum of the form $F_{\nu}\propto \nu^{\alpha}$. The subsequent quiescent level is consistent with the level expected\cite{bjfm11} from an early-type galaxy at $z\sim 0.5$. To estimate the likelihood that this transient could occur by chance we consider the results of radio imaging surveys (see Methods). By comparing to a recent survey with the VLA\cite{mooley} in the $2-4$~GHz band, we expect a $95\%$ ($99\%$) confidence upper limit of $<0.001$ ($<0.002$) such transients to occur in the ATCA observations of the FRB field, or equivalently an upper limit chance coincidence probability of $<0.1\%$ ($<0.2\%$). We find that the detection of a fading transient source is therefore sufficiently rare that we conclude that it is the afterglow from the FRB event.

We obtained optical observations of the field using Suprime-Cam on the 8.2-m Subaru Telescope on 2015 April 19 and April 20. From these images, we identify a source within the $\sim1$~arcsec positional uncertainty derived from the ATCA image (Figure 3). The source is seen to be a galaxy with a half-light radius of $1.4\pm0.2$~arcsec with a surface brightness profile well fit by de Vaucouleurs law (see Methods), thus consistent with being an elliptical galaxy. In addition to the Subaru $r^{\prime}$ and $i^{\prime}$ band photometry, we obtained $J$, $H$ and $K_{\mathrm{s}}$ band photometry with the Wide-field Infrared Camera on the Palomar 200'' telescope. The source is also detected in WISE W1 and W2 filters, enabling a 7-band spectral energy distribution fit (see Extended Data Figure 1) and the determination of a photometric redshift of $0.48 < z_{\mathrm{phot}} < 0.56$ ($68\%$ confidence). To confirm this redshift, we performed a 3-hour spectroscopic observation in good seeing conditions (FWHM $\sim 0.8$~arcsec) using FOCAS on Subaru on 2015 November 03. This yielded a spectrum consistent with a reddened elliptical galaxy at $z=0.492\pm 0.008$ (Figure 3). An earlier 1-hour observation, on a night which was not spectrophotometric, using DEIMOS on Keck, was also taken and found to be consistent with the Subaru result. 

Dispersion in the intergalactic medium (IGM) is related to the cosmic density of ionised baryons and the redshift\cite{ioka03,inoue04} according to the following expression:
\begin{equation}\label{eq:DM-z-baryons}
  \mathrm{DM}_{\mathrm{IGM}} = \frac{3 c H_0 \Omega_{\mathrm{IGM}}}{8\pi G m_{\mathrm{p}} }
  \int_{0}^{z_{\mathrm{FRB}}} \frac{(1+z^{\prime})f_{\mathrm{e}}(z^{\prime})dz^{\prime}}
      {[(1+z^{\prime})^3\Omega_{\mathrm{m}} + \Omega_{\mathrm{\Lambda}}]^{0.5}}\;.
\end{equation}
Here, we take $\mathrm{DM}_{\mathrm{IGM}}=\mathrm{DM}_{\mathrm{FRB}}-\mathrm{DM}_{\mathrm{MW}}-\mathrm{DM}_{\mathrm{halo}}-\mathrm{DM}_{\mathrm{host}}(1+z)^{-1}$ (see Methods). It is appropriate to account for a Milky Way halo contribution\cite{dgbb15} of $\mathrm{DM}_{\mathrm{halo}}=30\;\mathrm{cm}^{-3}\;\mathrm{pc}$ and we derive $\mathrm{DM}_{\mathrm{MW}}=189\;\mathrm{cm}^{-3}\;\mathrm{pc}$ from the NE2001 Galactic electron density model\cite{cl02}. An elliptical galaxy can be modelled\cite{xh15} with a modified version of NE2001 with an average rest-frame value of $\sim 37\;\mathrm{cm}^{-3}\;\mathrm{pc}$. The NE2001 components are uncertain at the $20\%$ level\cite{cl02}, and there is an additional uncertainty of $\sim 100\;\mathrm{cm}^{-3}\;\mathrm{pc}$ due to line-of-sight inhomogeneities in the IGM\cite{mcq14}. We therefore obtain:
\begin{equation}\label{eq:omega}
  \Omega_{\mathrm{IGM}} = \left(\frac{0.88}{f_{\mathrm{e}}}\right) 0.049 \pm 0.013 \;,
\end{equation}
where the ionisation factor $f_{\mathrm{e}}=1$ for $100\%$ ionised hydrogen, whereas allowing for a helium abundance of $25\%$, $f_{\mathrm{e}}=0.88$ is appropriate\cite{inoue04}. From fitting  $\mathrm{\Lambda}$CDM cosmological models to WMAP observations one derives\cite{wmap13} $\Omega_{\mathrm{baryons}}=0.046\pm 0.002$. Of these, $\sim 10\%$ are not ionised and/or in stellar interiors\cite{fp04} so that we expect to measure $\Omega_{\mathrm{IGM}}\approx 0.9\times\Omega_{\mathrm{baryons}} \approx 0.041\pm 0.002$. Thus, our measurement independently verifies the $\mathrm{\Lambda}$CDM model and the WMAP observations, and constitutes a direct measurement of the ionised material associated with all of the baryonic matter in the direction of the FRB, including the so-called ``missing'' baryons\cite{breg07}. Alternatively, if we take $\Omega_{\mathrm{IGM}}\equiv 0.041$, our measurements show that $\mathrm{DM}_{\mathrm{host}}$ is negligible.

FRB localisation allows us to correct a number of observable quantities that are corrupted by the unknown gain correction factor due to a lack of knowledge of the true position within the telescope beam. Accounting for the freqeuency-dependent beam response\cite{swb+96} we can derive a true spectral index for the FRB. We obtain $\alpha=+1.3 \pm 0.5$ for a fit centred at $1.382$~GHz. Similarly we derive a corrected flux density and fluence, and these, in combination with the redshift measurement, enable us to derive the distances, the energy released, the luminosity and other parameters (Table 1). 

In considering the nature of the progenitor we first consider the host galaxy. An upper limit to the star formation rate (SFR) can be determined from the upper limit $H_{\alpha}$ luminosity indicated by the Subaru spectrum (see Methods); the result is $\mathrm{SFR} \lesssim 0.2\;\mathrm{M}_{\odot}\;\mathrm{yr}^{-1}$. Such a low star formation rate implies, in the simplest interpretation, that FRB models directly related to recent star formation such as magnetar flares or blitzars are disfavoured. This problem might be avoided if either some residual star formation occurred in an otherwise `dead' galaxy, or if the FRB originated in one of the many satellite galaxies which are expected to surround an elliptical galaxy at this redshift, but which cannot be resolved in our observations. Failing these, the lack of star formation favours models such as compact merger events. This may be supported by the existence of the $\sim6$-day radio transient, which we interpret as the afterglow from the FRB. The presence of this short duration radio afterglow would not be expected from models involving giant pulses from pulsars, whereas supernovae and tidal disruption events would be expected to have a longer lasting afterglow\cite{pfk15}. The afterglow is consistent with the four short GRBs where radio afterglows have been detected\cite{fbma15}. The stellar mass of the galaxy, derived from the optical photometry and spectroscopy (see Methods), is $\approx 10^{11}\,M_\odot$, which is in accord with the masses of elliptical host galaxies of short GRBs\cite{berg14}. For a duration of 6 days for the afterglow we infer a brightness temperature of $\sim 10^{14}$~K, above the Compton cooling limit of $10^{12}$~K, implying a modest Doppler boosting factor. It is possible that the initial variability in brightness and spectral index is caused by scintillation, similar to that seen in GRB light-curves\cite{fkn+97}. If so, this indicates a very compact initial source size. 

Our conclusion that FRB~150418 is likely to be from a one-off event in an older stellar population may be at odds with the recent discovery\cite{mls+15} of FRB~110523. It is possible that there are two or more classes of FRB progenitor. These FRBs differ in their observed pulse widths and we speculate that this parameter might act as a discriminator between differing progenitors. FRB 150418 is not resolved (in time), but FRB 110523 appears to be (see Methods). This trend is seen in the published FRB population: some are clearly resolved, some are clearly not, although extra care needs to be taken in re-analysing this, as very slight errors in dispersion measure, as well as the effect of multi-path scattering, can make an FRB appear to have a longer intrinsic timescale.

Based on these facts, unresolved FRB~150418-like events might be asrcibed to cataclysmic events (such as short-GRBs), while FRB~110523-like events, where the intrinsic timescale is $\sim 1$~ms, could be associated to magnetar flares\cite{shri15}. In this case, the former class are unique events, whereas FRB~110523-like events could repeat, although the timescale for this is unclear. Answering these questions, and others such as whether or not some or all FRBs are standard candles, requires repeating the study we report here for a large number of FRBs, and continued monitoring of the known FRB fields.


\begin{addendum}
 \item The Parkes radio telescope and the Australia Telescope Compact
   Array are part of the Australia Telescope National Facility which
   is funded by the Commonwealth of Australia for operation as a
   National Facility managed by CSIRO. Parts of this research were
   conducted by the Australian Research Council Centre of Excellence
   for All-sky Astrophysics (CAASTRO) and utilised the gSTAR national
   facility at Swinburne University of Technology. Parts of this work
   are based on data collected at the Subaru Telescope which is
   operated by the National Astronomical Observatory of Japan, the
   Murchison Radio-astronomy Observatory operated by CSIRO, the GMRT
   which is run by the National Centre for Radio Astrophysics of the
   Tata Institute of Fundamental Research, the Sardinia Radio
   Telescope as part of scientific commissioning of the telescope, and
   the 100-m telescope of the MPIfR at Effelsberg. We acknowledge the
   Wajarri Yamatji people as the traditional owners of the MWA
   Observatory site.
 \item[Author Contributions] E.F.K. is PI of the SUPERB project,
   created SUPERB survey infrastructure at Parkes and Swinburne, led
   survey planning, formulated and wrote (with input from co-authors)
   the contents of this manuscript, performed the
   $\Omega_{\mathrm{IGM}}$ calculation, calculated the FRB spectral
   index, produced the FRB waterfall plot and the light curve
   plot. S.J. and S.B. performed ATCA observations and data
   analysis. S.J. and B.W.S. worked on radio light curve
   interpetation. S.B., N.D.R.B. and P.C. performed GMRT observations
   and data analysis. E.B. created survey infrastructure at Parkes and
   Swinburne and created the MWA shadowing
   infrastructure. Additionally E.F.K., S.J., S.B., E.B., N.D.R.B.,
   M.Bu, M.C., C.F., M.Kr., E.P., A.P., W.vS., M. Ba., S.B.-S. and
   R.P.E. all performed observations for the SUPERB survey at
   Parkes. A.J. created and maintained the Parkes and Swinburne
   hardware and software infrastructure and performed data management
   for the SUPERB project. M.Ba. additionally provided Parkes and
   Swinburne hardware. C.F. and M.Kr. also worked on the
   $\Omega_{\mathrm{IGM}}$ calculation. M.Kr. additionally performed
   FRB radio profile fitting. Polarisation analysis of the FRB signal
   was performed by M.C., E.P. and W.vS. W.vS. also produced the
   polarisation profile plot. E.P. additionally performed the
   \textit{Swift} analysis. Non-imaging radio followup was performed
   by M.Bu. A.P. and D.P. with the Sardinia Radio Telescope, by
   R.P.E. and M.Be. with the Effelsberg Radio Telescope, and by
   B.W.S., M.M. and C.B. at the Lovell Telescope at Jodrell
   Bank. T.To., M.H., H.F., T.H., T.M., Y.N., H.S., T.Te., N.T,
   S.Y. and N.Y. performed the Subaru observations. T.To., T.H.,
   N.T. and S.Y. additionally performed Subaru data analysis,
   determined the spectral redshift and created the optical profile
   plot. C.F., T.To., S.Y. and R.A. performed the optical profile
   fitting. J.C. performed data analysis on the Keck and Subaru data,
   also obtained the spectral redshift and produced the optical
   spectrum plot. J.J. performed the Palomar
   observations. M.Ka. performed the Keck observation. MWA
   observations were performed by N.D.R.B., D.K. S.J.T., A.W. and
   R.W. with data analysis by D.K. and S.J.T. D.K. additionally
   measured the photometric redshift and produced the RGB image and
   photo-z plots.
 \item[Author Information] Reprints and permissions information is
   available at www.nature.com/reprints. The authors declare no
   competing financial interests. Readers are welcome to comment on
   the online version of the paper. Correspondence and requests for
   materials should be addressed to E.F.K. (E.Keane@skatelescope.org) \\
\end{addendum}

\begin{table*}
  \begin{centering}
    \caption{Summary of FRB 150418 observed and derived
      properties. The peak flux density is a band average and a lower
      limit due to the intrinsic width not being resolved; similarly
      luminosity is a lower limit. The energy quoted is the product of
      the band-averaged fluence, the blueshifted effective bandwidth
      of the observations and the square of the luminosity distance.}
    \begin{tabular}{cc}

      \hline
      \hline
      Event time at $\nu_\mathrm{1382\;\mathrm{MHz}}$ & 2015 April 18 04:29:07.056 UTC \\[-2ex] 
      Event time at $\nu_\infty$ & 2015 April 18 04:29:05.370 UTC \\[-2ex] 
      Parkes beam number & 4, inner ring \\[-2ex] 
      Beam FWHM diameter & $14.1'$ \\[-2ex] 
      Beam Centre (RA, Dec) & (07:16:30.9, -19:02:24.4) \\[-2ex] 
      Beam Centre ($\ell$,$b$) & ($232.684^{\circ}$, $-3.261^{\circ}$) \\[-2ex] 
      Galaxy Position (RA, Dec), ATCA Epoch 1 & (07:16:35(3), -19:00:40(1)) \\[-2ex] 
      Galaxy Position ($\ell$,$b$), ATCA Epoch 1 & ($232.6654(1)^{\circ}$, $-3.2348(3)^{\circ}$) \\[-2ex] 
      Signal-to-noise ratio & 39 \\[-2ex] 
      Observed width, $W_{\mathrm{obs}}$ & $0.8\pm0.3$~ms \\[-2ex] 
      DM$_\mathrm{FRB}$ & $776.2(5)\;\mathrm{cm}^{-3}\;\mathrm{pc}$ \\[-2ex] 
      Dispersion index, $\beta$ & -2.00(1) \\[-2ex] 
      DM$_\mathrm{MW}$ & $188.5\;\mathrm{cm}^{-3}\;\mathrm{pc}$ \\[-2ex] 
      z$_{\mathrm{FRB}}$ & 0.492(8) \\[-2ex] 
      Peak flux density, $S_{1382\;\mathrm{MHz}}$ & $>2.2^{+0.6}_{-0.3}$~Jy (beam centre) \\[-2ex] 
      & $>2.4^{+0.5}_{-0.4}$~Jy (galaxy position) \\[-2ex] 
      Fluence, $\mathcal{F}_{1382\;\mathrm{MHz}}$ (Jy ms) &  $1.9^{+1.1}_{-0.8}$~Jy\;ms (beam centre) \\[-2ex] 
      & $2.0^{+1.2}_{-0.8}$~Jy\;ms (galaxy position) \\[-2ex] 
      Linear polarisation, $L/I$ & $8.5\pm1.5\%$ \\[-2ex] 
      Circular polarisation, $|V|/I$ & $<4.5\%$ ($3-\sigma$) \\[-2ex] 
      Rotation measure & $36\pm52\;\mathrm{rad}\;\mathrm{m}^{-2}$ \\[-2ex] 
      Spectral index, $\alpha$ & $\alpha > -3.0$ ($3-\sigma$, Parkes-MWA) \\[-2ex] 
      & $\alpha = +1.3 \pm 0.5$ (Parkes) \\[-2ex] 
      Comoving distance & $1.88$~Gpc \\[-2ex] 
      Luminosity distance & $2.81$~Gpc \\[-2ex] 
      Energy & $8^{+1}_{-5}\times 10^{38}$~ergs (galaxy position) \\[-2ex] 
      Luminosity & $>1.3 \times 10^{42}$~ergs/s (galaxy position) \\ 
      \hline

    \end{tabular}
  \end{centering}
\end{table*}


\newpage
\begin{figure}[H]
  \includegraphics[scale=0.5]{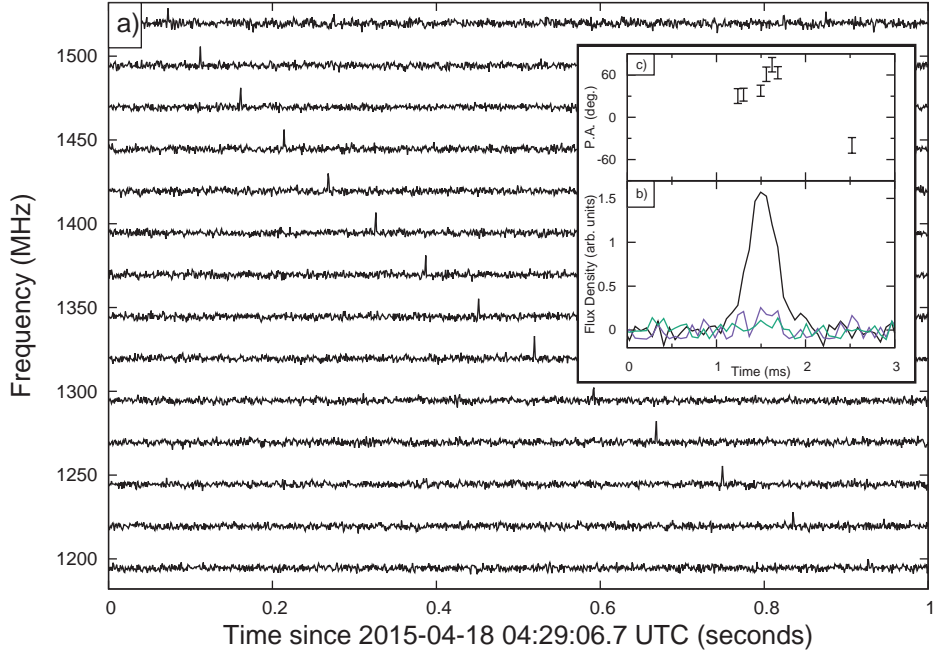}
  \caption{The FRB 150418 radio signal. Panel a shows a waterfall plot
    of the FRB signal with 15 frequency sub-bands across the Parkes
    observing bandwidth, showing the characteristic quadratic
    time-frequency sweep. To increase the signal-to-noise ratio the
    time resolution is reduced by a factor of 14 from the raw
    $64\;\upmu$s value. Panel b show the pulse profile of the FRB
    signal with the total intensity, linear and circular polarisation
    shown as black, purple and green lines respectively. Panel c shows
    the polarisation position angle with $1-\sigma$ error bars, for
    each $64$-$\upmu$s time sample where the linear polarisation was
    greater than twice the uncertainty in the linear polarisation.}
\end{figure}

\begin{figure}[H]
  \includegraphics[scale=0.5]{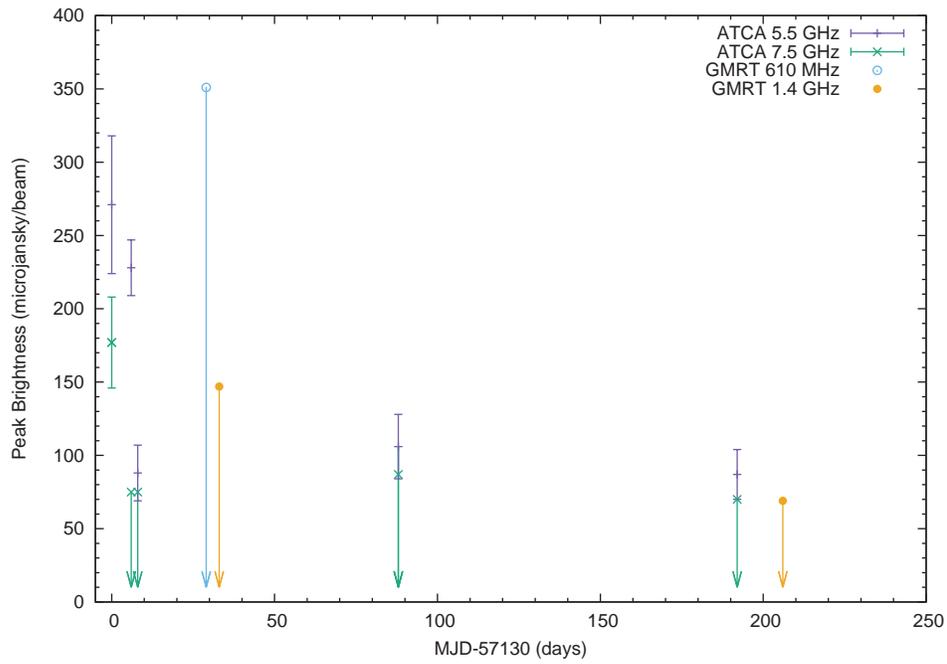}
  \caption{The FRB host galaxy radio light curve. Detections have
    $1-\sigma$ error bars, and $3-\sigma$ upper limits are indicated
    with arrows. The afterglow event appears to last $\sim 6$ days
    after which time the brightness settles at the quiescent level for
    the galaxy.}
\end{figure}

\begin{figure}[H]
  \includegraphics[scale=0.5]{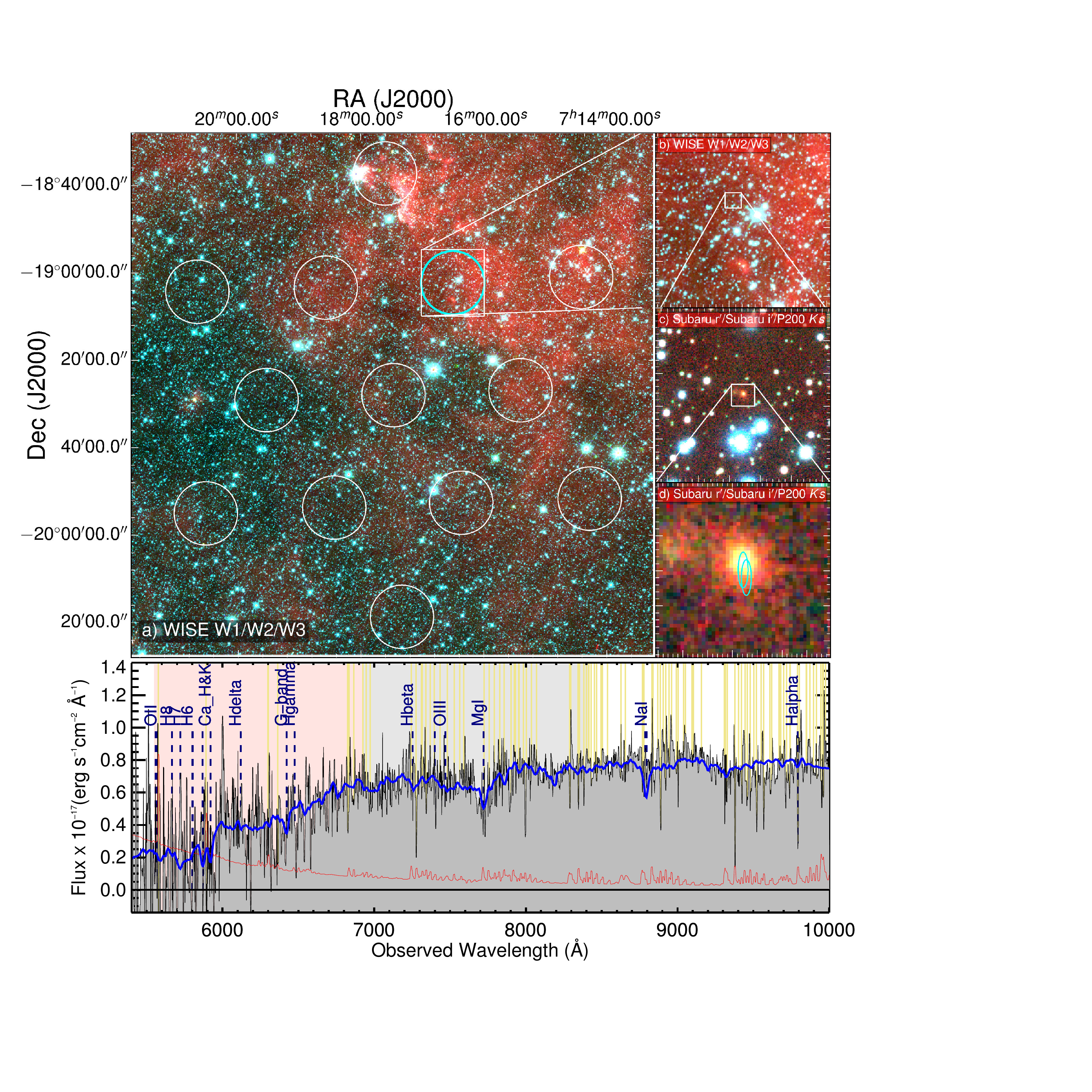}
  \vspace{-2cm}
  \caption{Optical analysis of the FRB host galaxy. Panel a shows a
    wide-field composite false-colour RGB image, overplotted with the
    half-power beam pattern of the Parkes multi-beam receiver. Panels
    b and c show successive zooms on the beam 4 region, and on the
    fading ATCA transient location respectively. Panel d is further
    zoomed in with the error ellipses for the ATCA transient, as
    derived from both the first and second epoch, overplotted. Panel e
    shows the Subaru FOCAS spectrum de-reddened with $E(B-V)=1.2$,
    smoothed by 5 pixels and fit to an elliptical galaxy template at
    $z=0.492$, denoted by the blue line. Common atomic transitions
    observed in galaxies are denoted by vertical dashed lines and
    yellow lines denote bright night sky lines. The Subaru
    $r^{\prime}$ and $i^{\prime}$ filter bandpasses are denoted by
    light red and grey background shading.}
\end{figure}

\newpage
\begin{methods}

\subsection{Dispersion Measure Contributions}
When radio signals propagate through a plasma, the travel time is longer than the light travel time \textit{in vacuo}. The additional delay depends on the radio wave frequency, $f$, and obeys a precise physical law of the form: $t_{\mathrm{DM}} = 4.149\;\mathrm{ms}\;(\mathrm{DM} /(\mathrm{cm}^{-3}\mathrm{pc}))(f/\mathrm{GHz})^{-2}$, where the dispersion measure (DM) is the integrated electron density along the line of sight. If travelling across cosmological distances there are several contributions to the observed DM --- from the host, the intergalactic medium (IGM) and the Milky Way. The quantity of interest is the IGM component as this can be used, in conjunction with redshift measurements, to perform a number of fundamental studies, e.g. detecting the missing baryons\cite{mcq14}, determining the dark energy equation of state\cite{zlw+14} and, if the signal is linearly polarised, measuring magnetic field strengths in the IGM\cite{zok+14}. For these applications the DM contributions other than $\mathrm{DM}_{\mathrm{IGM}}$ are essentially foregrounds which must be understood so that they can be removed. Here we examine each contribution in turn and note the uncertainties in each case; it is important to note that the precision in the observed DM value is high (see Table 1) and does not therefore contribute to the uncertainty in determining $\mathrm{DM}_{\mathrm{IGM}}$.

The Milky Way component consists of two parts, the first of which is that due to the interstellar medium. This is in principle known\cite{cl02} as the NE2001 model of the Galaxy's electron density can determine this to an accuracy of approximately $20\%$. The second Milky Way component is the contribution from the dark matter halo which is thought to exist, yet which is not included in the NE2001 model. We follow previous work\cite{dgbb15} which has calculated this term to be $30\;\mathrm{cm}^{-3}\;\mathrm{pc}$. It is unclear how to assign an uncertainty to this component so (considering the other components are dominant regardless) we take it to be zero. The host component is supressed by a factor of $(1+z)$. Its magnitude depends on both the nature of the progenitor and of the host galaxy. The observed FRB signal can be used to constrain the progenitor's local DM contribution --- a non-negligible contribution could imply a high density of electrons to be located close to the source. Such a high density configuration\cite{den14} could produce higher order dispersion terms (i.e. the pulse's frequency dispersion would deviate from quadratic), could result in plasma frequencies comparable to the emitted frequency (i.e. radiation would not escape), and could produce scattering in the pulse profile. None of these effects are observed for FRB~150418 implying that any local-progenitor DM component is negligible. The host galaxy contribution has been examined recently\cite{xh15} for spiral, dwarf and elliptical galaxies. This work considered modified versions of NE2001, with various sub-components of the model included/excluded as appropriate and suitable scalings to the $H_{\alpha}$ luminosity applied. For an elliptical galaxy, which is relevant in the case of FRB~150418, the average DM contribution over all inclination angles is $37\;\mathrm{cm}^{-3}\;\mathrm{pc}$ (this is the value before being supressed by the $(1+z)$ factor) and we use this as our estimate of the host contribution. As this is based upon NE2001 we assume that a $20\%$ uncertainty applies. In addition to the uncertainties mentioned already the IGM component itself is uncertain at the level of $\sim 100\;\mathrm{cm}^{-3}\;\mathrm{pc}$, due to inhomogeneities between different lines of sight through the IGM\cite{mcq14}.

\subsection{SUPERB}
SUPERB is a project ongoing at the 64-m Parkes radio telescope since April 2014 with goals of discovering FRBs and pulsars. The central frequency of the survey is 1.382~GHz, with a bandwidth of 400~MHz, of which $\sim 340$~MHz is typically usable. It uses optimised GPU codes for performing real-time radio frequency interference mitigation and searches for short duration radio bursts and pulsars in relativistic binary systems. In the real time search we use the following criteria to define candidate FRB events: (i) The DM of the burst must be at least $1.5$ times the expected maximum Milky Way contribution; (ii) The signal-to-noise ratio must be at least 10; (iii) The signal cannot be detected in more than 4 beams of the 13-beam receiver used for the SUPERB project at Parkes --- an event detected in more beams cannot originate from a boresight signal and therefore cannot be of a celestial origin; (iv) The width must be less than or equal to $8.192$~ms, i.e. $128$ times our native time sampling of $64\;\upmu$s; and (v) the number of independent events detected in a 4-second window centred on the event in question must be no greater than 5. The lag between the FRB signal hitting the dish and our software informing us of the detection\cite{pbb+15} is only $\sim 10$ seconds. We further search the data, offline, with more stringent interference rejection and covering corners of parameter space ignored for expediency during the real-time search. We note that since instigating this search system at Parkes no FRB has been missed by the real time search pipeline, including FRB 150418.

FRB 150418 was detected in beam 4 of the 21-cm multi-beam receiver. The FRB profile was fit simultaneously for time of arrival, dispersion measure, width, amplitude and dispersion index using 4 and 8 different sub-bands. The results were consistent with an unresolved pulse, where the width is purely given by the dispersion smearing across the $390.625$~kHz filterbank channels. Uncertainties were determined using CERN's MINUIT packages. The burst was found to have dispersion measure of $776.2(5)\;\mathrm{cm}^{-3}\;\mathrm{pc}$, and a dispersion index of $\beta = -2.00(1)$, where the dispersion delay is $\propto \nu^{\beta}$, consistent with propagation through a cold plasma.

The gain of beam 4 is well fit by a Gaussian\cite{swb+96} with FWHM of $14.1$~arcmin, so to derive corrected values for the flux density, fluence, etc. we boost the measured values by a factor of $\exp(\ln 2 (2\theta/\mathrm{FWHM})^2)$ where $\theta$ is the offset of the signal from the beam centre. The offset between the ATCA position determined from the first epoch (RA 07:16:34.557, Dec $-$19:00:39.954) and the centre of the Parkes beam (RA 07:16:30.9, Dec $-$19:02:24.4) is $\theta=1.944$~arcmin, yielding a boost factor of $1.054$. The observed peak flux density, if the FRB were at the centre of the beam, is $2.2$~Jy with a corresponding fluence of $1.9$~Jy\;ms; correcting these to the location of the host galaxy we estimate values of $2.4$~Jy and $2.0$~Jy\;ms. 

A calibration observation was taken 17 minutes post-burst and the polarisation calibration was performed using the PSRCHIVE software package\cite{hsm04}. Based on observations of PSR~J1644$-$4459, a bright polarised pulsar which we use to calibrate the off-axis response, taken three days prior to the FRB, we determine that the difference in the Jones matrix coefficients is not statistically different off boresight. Therefore the boresight calibration was used to determine the polarisation fraction of the pulse. This FRB is not seen to have a large linear polarisation, the rotation measure corrected linear polarisation is $L/I=8.5\pm1.5\%$ and the circular polarisation is consistent with zero. An additional systematic uncertainty exists in the leakage of total intensity to polarisation. Our PSR~J1644$-$4459 analysis provides an upper limit on the magnitude (but not orientation) of the leakage vector to be $<6\%$ of the total intensity, meaning that the true $L/I$ value may be either smaller or larger than quoted by up to this amount. Due to the low linear polarisation the rotation measure estimate is not very precise at $\mathrm{RM}=36\pm52\;\mathrm{rad}\;\mathrm{m}^{-2}$. As the RM is consistent with zero, examination of the IGM magnetic field strength is not possible with this FRB, although for completeness we note that the $3-\sigma$ upper limit on the electron weighted IGM magnetic field strength is $\sim 0.4\;\upmu$G for this line of sight. There is no evidence for a large host contribution to the RM for this FRB, although we note that an extremely large RM $\gtrsim 10^5\;\mathrm{rad}\;\mathrm{m}^{-2}$ would result in depolarisation within a single frequency channel meaning we are insensitive to such large values.

The MWA\cite{mwa13} was shadowing our Parkes observations but did not detect a counterpart. The resulting $3-\sigma$ fluence upper limit of $1050$~Jy\;ms at $185$~MHz gives us the first simultaneous multi-frequency observation of an FRB, and hence the first broadband limit on the spectral index. The spectral index limit from the Parkes and MWA data combined is $\alpha>-3.0$. Properties of the FRB are summarised below in Table 1.

\subsection{Followup Observations}
After the discovery of the FRB we triggered observations at numerous telescopes and performed a calibration observation at Parkes. We continued to observe with Parkes, obtaining 4.5 hours of observation over the course of the next 7.5 hours, in order to search for any repeat bursts. The MWA was shadowing during the discovery observation and continued to track the FRB position for another $\sim 7.5$ hours. The ATCA was on source 2 hours after the burst and also observed until T+7.5 hours, when the source set at both Parkes and ATCA. \textit{Swift} was on source 8 hours after the burst, and 10 hours after the burst, the Lovell Telescope continued the monitoring for 2.5 hours. On April 19 and 20 we obtained optical observations with Subaru, and on April 20 and 21 continued to search for repeated radio bursts with the Effelsberg, Sardinia and Parkes radio telescopes. The longer term followup campaign consisted of radio imaging (four further ATCA epochs, three GMRT epochs), high time resolution radio (with the Lovell Telescope), X-ray (one further \textit{Swift} epoch), optical photometry (with Palomar) and optical spectroscopy (with Keck and Subaru). We did not detect any followup bursts in our high time resolution radio followup (limiting flux densities in Extended Data Table 1), however regular emission at a much weaker level cannot be ruled out. The followup observations are summarised in Extended Data Table 1. We additionally note that no GRB was detected in temporal coincidence, or in the months before, by either \textit{Fermi} or \textit{Swift}. Furthermore, at a comoving (luminosity) distance of $1.8$~Gpc ($2.8$~Gpc), this galaxy is beyond the LIGO horizon for gravitational wave signatures from short GRBs\cite{ligo09}. 

\subsection{Imaging Transients}
The radio transient sky is not very well studied at frequencies of $5.5$ and $7.5$ GHz to the flux density levels relevant to this study. Additionally there are no archival data of the FRB field with which to compare our followup observations with. To estimate the likelihood of the $6$-day fading transient being detected by chance in our ATCA followup of the Parkes FRB field we first considered a previous ATCA study\cite{bhh+15}. This is the only such work performed on the same timescales and at the same observing frequency using the ATCA. In that work, which also covered a wider area of sky than our FRB followups and to a deeper level, no transient sources were discovered. A $95\%$ ($99\%$) confidence upper limit event rate of $<7.5\;\mathrm{deg}^{-2}$ ($<11.1\;\mathrm{deg}^{-2}$) with flux density of $>69\;\upmu$Jy/beam was obtained. Scaling this to obtain the expectation of a transient with flux density in excess of $200\;\upmu$Jy yields an upper limit event rate of $<1.5\;\mathrm{deg}^{-2}$ ($<2.2\;\mathrm{deg}^{-2}$). Considering the $0.04$-deg$^2$ field-of-view of a Parkes beam, which corresponds to the uncertainty in the FRB's position, the upper limit on the expected number of events in our followup observations is thereby $<0.06$ ($<0.09$). We can rephrase this by interpreting the upper limit number of expected events as the upper limit on $\lambda$, the Poisson rate parameter; then the probability of a chance temporal coincidence is $P(1;\lambda)=\lambda \exp(-\lambda)$. This yields an upper limit probability of $<6\%$ ($<8\%$).

One can also obtain estimates of the false-positive rate from considering studies\cite{bhwp10,ofb+11,fko+12,cbw13,gop+06} at other telescopes. As other studies are not ideally matched in terms of observing frequency and sensitivity, one must scale the findings appropriately. For example for studies performed at different observing frequencies we must scale the flux densities by a spectral index; we adopt the spectral index of our $6$-day transient as measured in the first epoch of our followup. Additionally, sensitivity levels must be scaled to the $200\;\upmu$Jy level; for this operation we adopt the standard $N\propto S^{-3/2}$ scaling. With this approach we consider a recent deep VLA study\cite{mooley} which operated between $2$ and $4$ GHz. Applying the appropriate scaling this study yields an expected number of events in our followup observations of $<0.001$ ($<0.002$) at $95\%$ ($99\%$) confidence. The equivalent upper limit chance temporal coincidence probability is $<0.1\%$ ($<0.2\%$). This result is more constraining than the ATCA-derived numbers by a factor of at least $60$. From this assessment we deem it statistically unlikely that we would have detected, by chance, this fading negative spectral index radio source at this location and time, resulting in our interpetation that this source is likely to be associated with the FRB.

Ideally we might expand upon this calculation by estimating probabilities of chance coincidence in each independent wave-band (radio, optical and X-ray) in our follow-up campaign, and then compute a joint probability of a transient occuring in any of the bands. It is unclear what the appropriate statistics are for the X-ray and optical bands, but if we take the observed GRB and supernova rates as indicative, we find that the upper limit expectation in these wave bands is much smaller than in the radio. Deeper all-sky radio transient surveys are therefore the key to tightening constraints on transient associations like that reported here.

\subsection{Optical Analysis}
The $i^{\prime}$ optical profile of the Galaxy was fit with a Sersic function of the form $I(r)\propto \exp(-k R^{1/n})$. The best fit parameters (see Extended Data Figure 2) are a Sersic index of $n=3.6\pm0.5$, consistent with the $n=4$ value seen in elliptical galaxies. The half-light radius is $R_{\mathrm{e}}=6.9\pm1.2$~pixels, or $R_{\mathrm{e}}=1.4\pm0.2$~arcsec. At $z=0.492$ the angular diameter distance is $1.62$~Gpc so that this implies $R_{\mathrm{e}}=10.9\pm1.8$~kpc as the physical half-light radius of the galaxy. The profile fitting also yields an estimate of the major to minor axes ratio for the galaxy of $b/a=0.68\pm0.03$. We note that there are additional systematic uncertainties involved in the fit of the Sersic index, depending on the exact method of sky subtraction employed; our analysis suggests that the systematic errors are probably of equal magntude to the statistical errors quoted above.

We obtained the following photometry of the FRB host galaxy: with Subaru Suprime-Cam\cite{mks+02} we determined AB magnitudes of $23.45(16)$ and $22.07(31)$, for $r^{\prime}$ and $i^{\prime}$ bands respectively. Between the two Subaru epochs no variability is seen --- a subtraction of the epochs yields an upper limit on any variation of $25.2$ and $24.7$ magnitudes ($5-\sigma$). The galaxy is also detected in two WISE filters with Vega magnitudes of $15.204(0.044)$ and $15.050(082)$, for the $W_1$ and $W_2$ bands respectively. With the Palomar 200'' (P200) telescope we obtained further information, obtaining Vega magnitudes of $18.92(10)$, $17.55(25)$ and $16.51(05)$ for $J$, $H$ and $K_{\mathrm{s}}$ bands respectively. We corrected the observed photometry for the Milky Way extinction using $A_{\mathrm{V}}=3.7\,$mag and standard extinction coefficients\cite{dib+14}, and converted those magnitudes into flux densities using the established zero-points\cite{cwm03,jcm+11}. We fit for the photometric redshift using the 2015~November version of \texttt{EAZY}\cite{bdc08}, finding 68\% confidence limits $0.48<z_{\rm phot} < 0.56$. This was robust to different choice of galaxy template, with good overall fits ($\chi^2$ of $5-8$). We then fit the SED of the host galaxy using \texttt{MAGPHYS}\cite{cce08}. We were able to achieve an acceptable fit (see Extended Data Figure 1) to all of the photometry with a model for a passive (SFR$\lesssim 0.2\;\mathrm{M}_{\odot}\;\mathrm{yr}^{-1}$), massive (stellar mass $\approx 10^{11}\,M_\odot$) galaxy with a modest amount of dust (in-host extinction $A_{\mathrm{V}}$ in the range $0-4$~mag). The exact fit was rather degenerate because of our limited wavelength coverage, with only the stellar mass well-determined. Future observations at shorter wavelengths should be able to determine more robust properties. 

The spectrum which confirmed the redshift was obtained on 2015 November 2 in a 3-hour observation using FOCAS on the Subaru telescope. An earlier attempt to obtain a spectroscopic redshift on 2015 October 21 using DEIMOS on Keck in poorer sky conditions had proven difficult to calibrate and resulted in an imprecise redshift estimate, no better than the photometric estimate. In that case difficulties in calibration are compounded by the spectrum's lack of any significant emission lines. The well calibrated Subaru spectrum is found to be consistent with a reddened $z = 0.492(8)$ early type galaxy with $E(B-V) = 1.2 \pm 0.1$ (Figure 3), noting that $r^{\prime}$ and $i^{\prime}$ are approximations to restframe $B$ and $V$ filters. This implies absolute magnitudes of $M_{\mathrm{B}}\approx -21.6$ and $M_{\mathrm{V}}\approx -22.1$. As the galaxy is elliptical we can apply the Faber-Jackson relation\cite{fj76} to estimate the velocity dispersion of $\sim 230$~km/s. From the Virial Theorem and the observed half-light radius we can thus estimate the stellar (and total) mass\cite{bks+01} to be $\sim 10^{11}\;\mathrm{M}_{\odot}$ (and $\sim 2\times 10^{12}\;\mathrm{M}_{\odot}$). An upper limit to the $H_{\alpha}$ luminosity of $2.6\times10^{40}$~erg/s ($3-\sigma$) can be derived from the optical spectrum, and from this we can, in the standard way\cite{ken98} derive a star formation rate of $\mathrm{SFR} \lesssim 0.2\;\mathrm{M}_{\odot}\;\mathrm{yr}^{-1}$.

The very faint companion galaxy ($r^{\prime}=24.22(16)$~mag, $i^{\prime}=23.22(31)$~mag) visible to the North-East is not inconsistent with a galaxy at the same redshift. If confirmed the two galaxies may be in the process of merging. The absolute $K$ band magnitude of the galaxy is $-25.7$~mag (Vega). The radio continuum level can be estimated from this\cite{bjfm11} and is consistent with the quiescent level of the galaxy after the $\sim 6$-day fading event. This shows that the background level seen is not surprising for an early type galaxy, and implies the FRB afterglow had already faded below this level by the 3rd ATCA epoch.

\subsection{FRB~110523} 
After the initial submission of this manuscript a study was published\cite{mls+15} announcing the discovery of FRB~110523, which we have compared and contrasted with FRB 150418 in the main text. There are now strong indications that there are two or more FRB progenitors, and we speculate that the observed pulse width may act as a useful discriminator between these. To this end we consider the time sampling and frequency resolution of the study that discovered FRB 110523, as well as its DM, in an effort to see if the pulse was resolved or not in the GBT observations. We would expect an unresolved FRB to have an observed width of no more than 1.08, 1.14 and 1.26~ms in the highest, central and lowest frequency channel respectively, in the Green Bank study. This is inconsistent with the reported observed width (after the effects of scattering had been removed) at the $3$ to $4-\sigma$ level indicating that FRB 110523 appears to be resolved and therefore to have an intrinsic timescale of $\sim 1$~ms.

\end{methods}



\newpage
\renewcommand{\tablename}{Extended Data Table}
\setcounter{table}{0}

\newpage
\begin{table*}
  \begin{centering}
    \caption{\small{Summary of followup observations of FRB
        150418. The start times, telescope name, observing band and
        observing duration are listed. In the fifth column time
        domain, radio imaging, photon counting, photometric and
        spectroscopic observations are denoted by TD, RI, PC, Ph and
        Sp respectively, and detections (of either the FRB, fading
        transient or host galaxy) are denoted with a an asterisk. The
        final column gives the detection level, or 3-$\sigma$ upper
        limits in the case of non-detections.}}
    \begin{tabular}{llllll}

      \hline
      \hline
      Time (UTC) & Telescope & Band & Tobs (s) & Mode & Level/limit \\
      \hline
      2015-04-18-04:21:15 & Parkes & 1.4 GHz     & 561          &  TD* & $2.2$~Jy  \\[-3ex]
      Shadowing           & MWA    & 185 MHz     & 27000        &  TD  & $<1050$~Jy\;ms\\[-3ex]
      2015-04-18-04:31:08 & Parkes & 1.4 GHz     & 465          &  TD  & $<0.17(W/0.9\;\mathrm{ms})^{-0.5}$~Jy \\[-3ex]
      2015-04-18-05:04:35 & Parkes & 1.4 GHz     & 1181         &  TD  & $<0.17(W/0.9\;\mathrm{ms})^{-0.5}$~Jy \\[-3ex]
      2015-04-18-06:30:15 & ATCA   & 5.5,7.5 GHz & 19800        &  RI* & $0.27(5)$~mJy/beam, $0.18(3)$~mJy/beam \\[-3ex]
      2015-04-18-07:46:27 & Parkes & 1.4 GHz     & 3618         &  TD  & $<0.17(W/0.9\;\mathrm{ms})^{-0.5}$~Jy \\[-3ex]
      2015-04-18-08:47:28 & Parkes & 1.4 GHz     & 3618         &  TD  & $<0.17(W/0.9\;\mathrm{ms})^{-0.5}$~Jy \\[-3ex]
      2015-04-18-09:48:09 & Parkes & 1.4 GHz     & 3617         &  TD  & $<0.17(W/0.9\;\mathrm{ms})^{-0.5}$~Jy \\[-3ex]
      2015-04-18-10:48:59 & Parkes & 1.4 GHz     & 3617         &  TD  & $<0.17(W/0.9\;\mathrm{ms})^{-0.5}$~Jy \\[-3ex]
      2015-04-18-11:49:37 & Parkes & 1.4 GHz     & 758          &  TD  & $<0.17(W/0.9\;\mathrm{ms})^{-0.5}$~Jy \\[-3ex]
      2015-04-18-12:20:57 & \textit{Swift} & X-ray & 3976       &  PC  & $<7.1\times10^{-14}$ erg/cm$^2$/s\\[-3ex]
      2015-04-18-14:22:52 & Lovell & 1.4 GHz     & 7200         &  TD  & $<0.11(W/0.9\;\mathrm{ms})^{-0.5}$~Jy \\[-3ex]
      2015-04-19-06:06:19 & Subaru & $i^{\prime}$ & 600         &  Ph* & 22.06(31) mag (AB) \\[-3ex]
      2015-04-19-06:37:12 & Subaru & $r^{\prime}$ & 900         &  Ph* & 23.33(16) mag (AB) \\[-3ex]
      2015-04-20-05:50:27 & Subaru & $i^{\prime}$ & 1200        &  Ph* & 22.08(31) mag (AB) \\[-3ex]
      2015-04-20-06:30:53 & Subaru & $r^{\prime}$ & 1200        &  Ph* & 23.59(16) mag (AB) \\[-3ex]
      2015-04-20-15:49:09 & Effelsberg & 1.4 GHz & 8300         &  TD  & $<0.11(W/0.9\;\mathrm{ms})^{-0.5}$~Jy \\[-3ex]
      2015-04-21-06:40:42 & Parkes & 1.4 GHz     & 3600         &  TD  & $<0.17(W/0.9\;\mathrm{ms})^{-0.5}$~Jy \\[-3ex]
      2015-04-21-17:21:40 & SRT    & 1.4 GHz     & 3600         &  TD  & $<0.61(W/0.9\;\mathrm{ms})^{-0.5}$~Jy \\[-3ex]
      2015-04-24-02:44:15 & ATCA   & 5.5,7.5 GHz & 72900        &  RI* & $0.23(2)$~mJy/beam, $<0.08$~mJy/beam \\[-3ex]
      2015-04-26-01:45:05 & ATCA   & 5.5,7.5 GHz & 74700        &  RI* & $0.09(2)$~mJy/beam, $<0.08$~mJy/beam \\[-3ex]
      2015-05-07-03:18:42 & \textit{Swift} & X-ray & 2908       &  PC  & $<9.3\times10^{-14}$ erg/cm$^2$/s\\[-3ex]
      2015-05-18-12:30:00 & GMRT   & 0.61 GHz    & 7200         &  RI  & $<0.35$~mJy/beam \\[-3ex]
      2015-05-22-12:42:00 & GMRT   & 1.4 GHz     & 7140         &  RI  & $<0.15$~mJy/beam \\[-3ex]
      2015-06-04-21:12:15 & ATCA   & 5.5,7.5 GHz & 26700        &  RI* & $0.11(2)$~mJy/beam, $<0.09$~mJy/beam \\[-3ex]
      2015-10-15-05:32:23 & Lovell & 1.4 GHz     & 7200         &  TD  & $<0.14(W/0.9\;\mathrm{ms})^{1/2}$~Jy\;ms \\[-3ex]
      2015-10-21-00:14:15 & Keck   & OIR         & 3600         &  Sp* & see Subaru observation\\[-3ex]
      2015-10-27-14:09:35 & ATCA   & 5.5,7.5 GHz & 30600        &  RI* & $0.09(2)$~mJy/beam, $<0.07$~mJy/beam \\[-3ex]
      2015-10-31-11:15:54 & P200   & $J$         & 1080         &  Ph* & 18.92(10) mag (Vega) \\[-3ex]
      2015-10-31-11:38:31 & P200   & $H$         & 450          &  Ph* & 17.55(25) mag (Vega) \\[-3ex]
      2015-10-31-11:51:15 & P200   & $K_{\mathrm{s}}$ & 810     &  Ph* & 16.51(5) mag (Vega) \\[-3ex]
      2015-11-03-11:57:41 & Subaru & OIR         & 10800        &  Sp* & see Figure 3\\[-3ex]
      2015-11-11-18:40:00 & GMRT   & 1.4 GHz     & 22020        &  RI  & $<0.07$~mJy/beam \\
      \hline

    \end{tabular}
  \end{centering}
\end{table*}


\renewcommand{\figurename}{Extended Data Figure}
\setcounter{figure}{0}

\begin{figure}
  \includegraphics[scale=0.2]{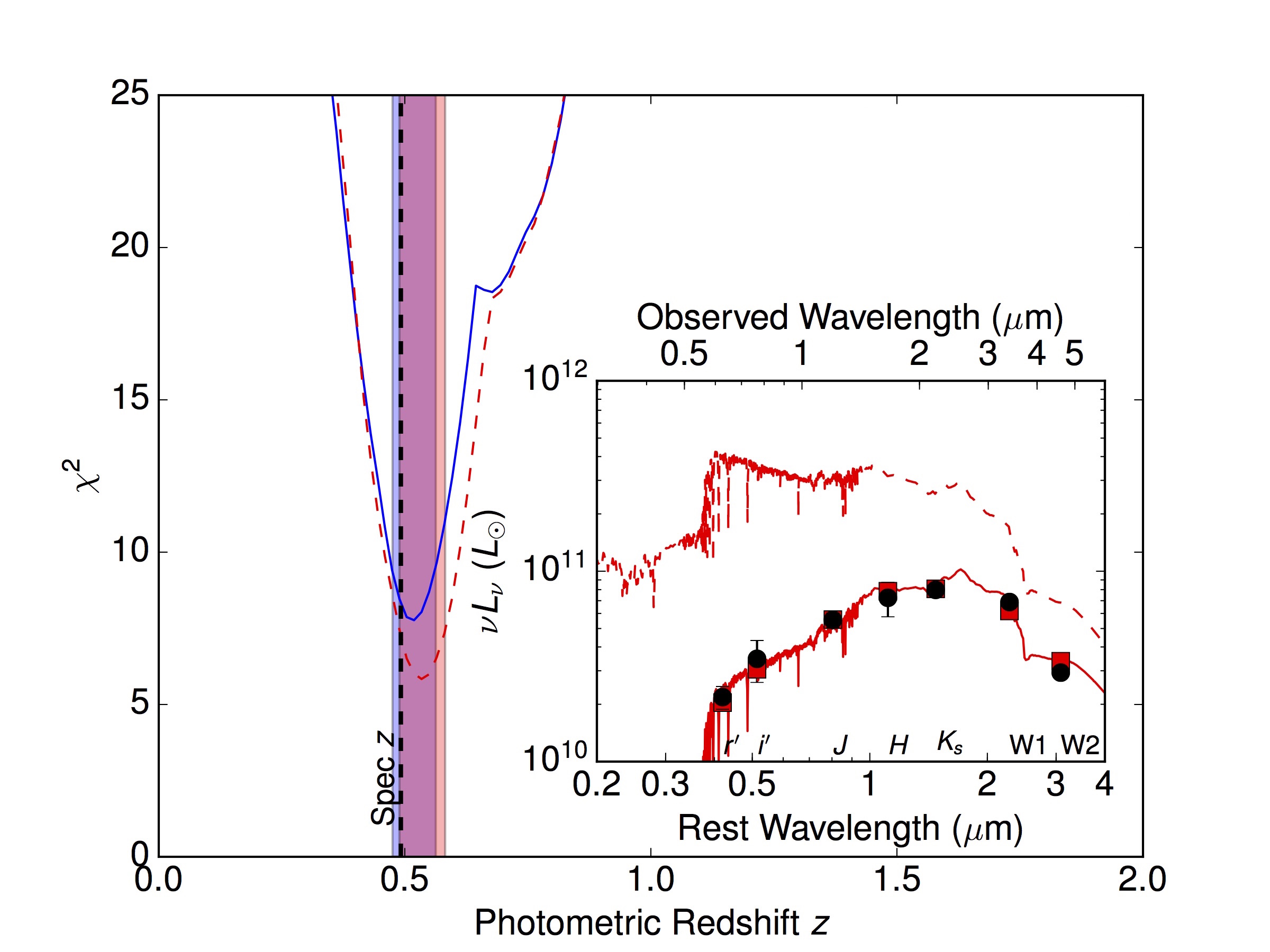}
  \caption{The Photometric Redshift of the FRB host galaxy. A $\chi^2$
    fit of the redshift of the galaxy based on the spectral energy
    distribution is shown. The photometric redshift determined from
    this is $0.48 < z_{\mathrm{phot}} < 0.56$ (68\% confidence,
    denoted by the shaded region). The inset shows the spectral energy
    distribution fit with the 7 photometric estimates overplotted with
    $1-\sigma$ error bars.}
\end{figure}

\begin{figure}
  \vspace{-5cm}
  \includegraphics[scale=0.2,trim = 0mm 60mm 0mm 60mm, clip]{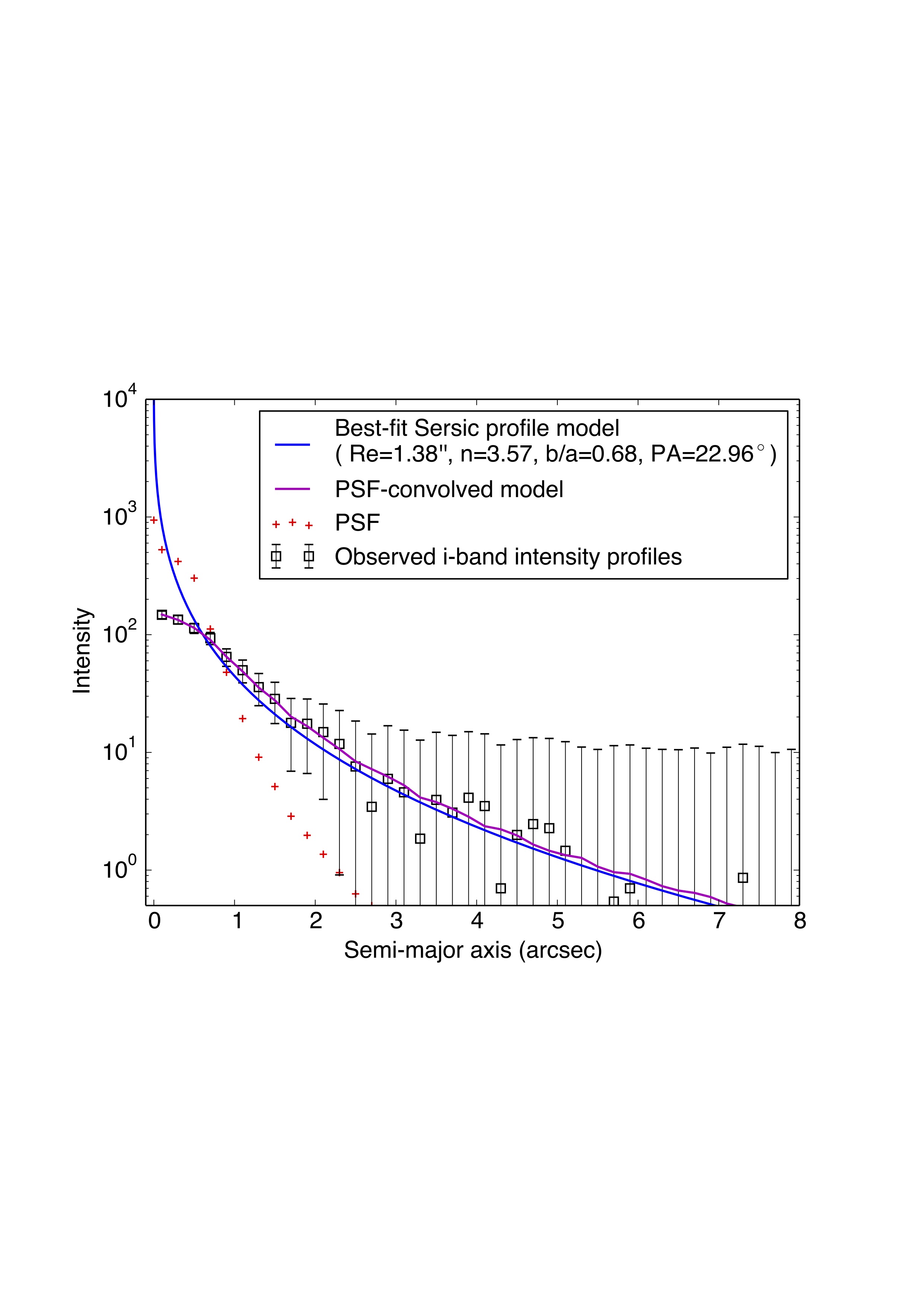}
  \vspace{-5cm}
  \caption{The optical surface brightness profile of the FRB host
    galaxy. The surface brightness profile of the galaxy in the Subaru
    $i^{\prime}$ band image was fit to an ellipsoidal Sersic
    function. The model profiles and data are shown as the flux along
    an ellipse as a function of semi-major axis. The image point
    spread function profile is also shown as a function of radius.}
\end{figure}


\newpage
\begin{thebibliography}{10}
\newcounter{firstbib}
\expandafter\ifx\csname url\endcsname\relax
  \def\url#1{\texttt{#1}}\fi
\expandafter\ifx\csname urlprefix\endcsname\relax\def\urlprefix{URL }\fi
\providecommand{\bibinfo}[2]{#2}
\providecommand{\eprint}[2][]{\url{#2}}

\bibitem{lbm+07}
\bibinfo{author}{{Lorimer}, D.~R.}, \bibinfo{author}{{Bailes}, M.},
  \bibinfo{author}{{McLaughlin}, M.~A.}, \bibinfo{author}{{Narkevic}, D.~J.} \&
  \bibinfo{author}{{Crawford}, F.}
\newblock \bibinfo{title}{{A Bright Millisecond Radio Burst of Extragalactic
  Origin}}.
\newblock \emph{\bibinfo{journal}{Science}} \textbf{\bibinfo{volume}{318}},
  \bibinfo{pages}{777--780} (\bibinfo{year}{2007}).

\bibitem{kskl12}
\bibinfo{author}{{Keane}, E.~F.}, \bibinfo{author}{{Stappers}, B.~W.},
  \bibinfo{author}{{Kramer}, M.} \& \bibinfo{author}{{Lyne}, A.~G.}
\newblock \bibinfo{title}{{On the origin of a highly dispersed coherent radio
  burst}}.
\newblock \emph{\bibinfo{journal}{Mon. Not. R. Astron. Soc.}} \textbf{\bibinfo{volume}{425}},
  \bibinfo{pages}{L71--L75} (\bibinfo{year}{2012}).

\bibitem{tsb+13}
\bibinfo{author}{{Thornton}, D.} \emph{et~al.}
\newblock \bibinfo{title}{{A Population of Fast Radio Bursts at Cosmological
  Distances}}.
\newblock \emph{\bibinfo{journal}{Science}} \textbf{\bibinfo{volume}{341}},
  \bibinfo{pages}{53--56} (\bibinfo{year}{2013}).

\bibitem{sch+14}
\bibinfo{author}{{Spitler}, L.~G.} \emph{et~al.}
\newblock \bibinfo{title}{{Fast Radio Burst Discovered in the Arecibo Pulsar
  ALFA Survey}}.
\newblock \emph{\bibinfo{journal}{Astrophys. J.}} \textbf{\bibinfo{volume}{790}},
  \bibinfo{pages}{101--110} (\bibinfo{year}{2014}).

\bibitem{bb14}
\bibinfo{author}{{Burke-Spolaor}, S.} \& \bibinfo{author}{{Bannister}, K.~W.}
\newblock \bibinfo{title}{{The Galactic Position Dependence of Fast Radio
  Bursts and the Discovery of FRB 011025}}.
\newblock \emph{\bibinfo{journal}{Astrophys. J.,}} \textbf{\bibinfo{volume}{792}},
 \bibinfo{pages}{19--26} (\bibinfo{year}{2014}).

\bibitem{rsj15}
\bibinfo{author}{{Ravi}, V.}, \bibinfo{author}{{Shannon}, R.~M.} \&
  \bibinfo{author}{{Jameson}, A.}
\newblock \bibinfo{title}{{A Fast Radio Burst in the Direction of the Carina
  Dwarf Spheroidal Galaxy}}.
\newblock \emph{\bibinfo{journal}{Astrophys. J.}} \textbf{\bibinfo{volume}{799}},
  \bibinfo{pages}{L5--L10} (\bibinfo{year}{2015}).

\bibitem{pbb+15}
\bibinfo{author}{{Petroff}, E.} \emph{et~al.}
\newblock \bibinfo{title}{{A real-time fast radio burst: polarization detection
  and multiwavelength follow-up}}.
\newblock \emph{\bibinfo{journal}{Mon. Not. R. Astron. Soc.}} \textbf{\bibinfo{volume}{447}},
  \bibinfo{pages}{246--255} (\bibinfo{year}{2015}).

\bibitem{mls+15}
\bibinfo{author}{{Masui}, K.} \emph{et~al.}
\newblock \bibinfo{title}{{Dense magnetized plasma associated with a fast radio
  burst}}.
\newblock \emph{\bibinfo{journal}{Nature}} \textbf{\bibinfo{volume}{528}},
  \bibinfo{pages}{523--525} (\bibinfo{year}{2015}).

\bibitem{cha+16}
\bibinfo{author}{{Champion}, D.} \emph{et~al.}
\newblock \bibinfo{title}{{Five new Fast Radio Bursts from the HTRU high
  latitude survey: rst evidence for two-component bursts}}.
\newblock \emph{\bibinfo{journal}{Mon. Not. R. Astron. Soc., submitted}}  (\bibinfo{year}{2015}).
\newblock \eprint{1511.07746}.

\bibitem{mcq14}
\bibinfo{author}{{McQuinn}, M.}
\newblock \bibinfo{title}{{Locating the ''Missing'' Baryons with Extragalactic
  Dispersion Measure Estimates}}.
\newblock \emph{\bibinfo{journal}{Astrophys. J.}} \textbf{\bibinfo{volume}{780}},
  \bibinfo{pages}{L33--L38} (\bibinfo{year}{2014}).

\bibitem{zlw+14}
\bibinfo{author}{{Zhou}, B.}, \bibinfo{author}{{Li}, X.},
  \bibinfo{author}{{Wang}, T.}, \bibinfo{author}{{Fan}, Y.-Z.} \&
  \bibinfo{author}{{Wei}, D.-M.}
\newblock \bibinfo{title}{{Fast radio bursts as a cosmic probe?}}
\newblock \emph{\bibinfo{journal}{PhysRevD}} \textbf{\bibinfo{volume}{89}},
  \bibinfo{pages}{107303} (\bibinfo{year}{2014}).

\bibitem{wmap13}
\bibinfo{author}{{Hinshaw}, G.} \emph{et~al.}
\newblock \bibinfo{title}{{Nine-year Wilkinson Microwave Anisotropy Probe
  (WMAP) Observations: Cosmological Parameter Results}}.
\newblock \emph{\bibinfo{journal}{Astrophys. J. Supp.}} \textbf{\bibinfo{volume}{208}},
  \bibinfo{pages}{19} (\bibinfo{year}{2013}).

\bibitem{cf12}
\bibinfo{author}{{Chandra}, P.} \& \bibinfo{author}{{Frail}, D.~A.}
\newblock \bibinfo{title}{{A Radio-selected Sample of Gamma-Ray Burst
  Afterglows}}.
\newblock \emph{\bibinfo{journal}{Astrophys. J.}} \textbf{\bibinfo{volume}{746}},
  \bibinfo{pages}{156} (\bibinfo{year}{2012}).

\bibitem{cl02}
\bibinfo{author}{{Cordes}, J.~M.} \& \bibinfo{author}{{Lazio}, T.~J.~W.}
\newblock \bibinfo{title}{{NE2001. I. A New Model for the Galactic Distribution
  of Free Electrons and its Fluctuations}}  (\bibinfo{year}{2002}).
\newblock \bibinfo{note}{Preprint (arXiv:astro-ph/0207156)}.

\bibitem{bcc+04}
\bibinfo{author}{{Bhat}, N.~D.~R.}, \bibinfo{author}{{Cordes}, J.~M.},
  \bibinfo{author}{{Camilo}, F.}, \bibinfo{author}{{Nice}, D.~J.} \&
  \bibinfo{author}{{Lorimer}, D.~R.}
\newblock \bibinfo{title}{{Multifrequency Observations of Radio Pulse
  Broadening and Constraints on Interstellar Electron Density Microstructure}}.
\newblock \emph{\bibinfo{journal}{Astrophys. J.}}
  \textbf{\bibinfo{volume}{605}}, \bibinfo{pages}{759--783}
  (\bibinfo{year}{2004}).

\bibitem{bhh+15}
\bibinfo{author}{{Bell}, M.~E.} \emph{et~al.}
\newblock \bibinfo{title}{{A search for variable and transient radio sources in
  the extended Chandra Deep Field South at 5.5 GHz}}.
\newblock \emph{\bibinfo{journal}{Mon. Not. R. Astron. Soc.}} \textbf{\bibinfo{volume}{450}},
  \bibinfo{pages}{4221--4232} (\bibinfo{year}{2015}).

\bibitem{bjfm11}
\bibinfo{author}{{Brown}, M.~J.~I.}, \bibinfo{author}{{Jannuzi}, B.~T.},
  \bibinfo{author}{{Floyd}, D.~J.~E.} \& \bibinfo{author}{{Mould}, J.~R.}
\newblock \bibinfo{title}{{The Ubiquitous Radio Continuum Emission from the
  Most Massive Early-type Galaxies}}.
\newblock \emph{\bibinfo{journal}{Astrophys. J.}} \textbf{\bibinfo{volume}{731}},
  \bibinfo{pages}{L41} (\bibinfo{year}{2011}).

\bibitem{mooley}
\bibinfo{author}{{Mooley}, K.~P.} \emph{et~al.}
\newblock \bibinfo{title}{{The Caltech-NRAO Stripe 82 Survey (CNSS) Paper I: The Pilot Radio Transient Survey In 50 deg$^2$}}.
\newblock \emph{\bibinfo{journal}{Astrophys. J.}} in press,
  (\bibinfo{year}{2016}).
\newblock \bibinfo{note}{arXiv:astro-ph/1601.01693}.

\bibitem{ioka03}
\bibinfo{author}{{Ioka}, K.}
\newblock \bibinfo{title}{{The Cosmic Dispersion Measure from Gamma-Ray Burst
  Afterglows: Probing the Reionization History and the Burst Environment}}.
\newblock \emph{\bibinfo{journal}{Astrophys. J.}} \textbf{\bibinfo{volume}{598}},
  \bibinfo{pages}{L79--L82} (\bibinfo{year}{2003}).

\bibitem{inoue04}
\bibinfo{author}{{Inoue}, S.}
\newblock \bibinfo{title}{{Probing the cosmic reionization history and local
  environment of gamma-ray bursts through radio dispersion}}.
\newblock \emph{\bibinfo{journal}{Mon. Not. R. Astron. Soc.}} \textbf{\bibinfo{volume}{348}},
  \bibinfo{pages}{999--1008} (\bibinfo{year}{2004}).

\bibitem{dgbb15}
\bibinfo{author}{{Dolag}, K.}, \bibinfo{author}{{Gaensler}, B.~M.},
  \bibinfo{author}{{Beck}, A.~M.} \& \bibinfo{author}{{Beck}, M.~C.}
\newblock \bibinfo{title}{{Constraints on the distribution and energetics of fast radio bursts using cosmological hydrodynamic simulations}}.
\newblock \emph{\bibinfo{journal}{Mon. Not. R. Astron. Soc.}}
  \textbf{\bibinfo{volume}{451}}, \bibinfo{pages}{4277--4289} (\bibinfo{year}{2015}).

\bibitem{xh15}
\bibinfo{author}{{Xu}, J.} \& \bibinfo{author}{{Han}, J.~L.}
\newblock \bibinfo{title}{{Extragalactic dispersion measures of fast radio
  bursts}}.
\newblock \emph{\bibinfo{journal}{Research in Astronomy and Astrophysics}}
  \textbf{\bibinfo{volume}{15}}, \bibinfo{pages}{1629} (\bibinfo{year}{2015}).

\bibitem{fp04}
\bibinfo{author}{{Fukugita}, M.} \& \bibinfo{author}{{Peebles}, P.~J.~E.}
\newblock \bibinfo{title}{{The Cosmic Energy Inventory}}.
\newblock \emph{\bibinfo{journal}{Astrophys. J.}} \textbf{\bibinfo{volume}{616}},
  \bibinfo{pages}{643--668} (\bibinfo{year}{2004}).

\bibitem{breg07}
\bibinfo{author}{{Bregman}, J.~N.}
\newblock \bibinfo{title}{{The Search for the Missing Baryons at Low
  Redshift}}.
\newblock \emph{\bibinfo{journal}{ARAA}} \textbf{\bibinfo{volume}{45}},
  \bibinfo{pages}{221--259} (\bibinfo{year}{2007}).

\bibitem{swb+96}
\bibinfo{author}{Staveley-Smith, L.} \emph{et~al.}
\newblock \bibinfo{title}{The {P}arkes 21 cm multibeam receiver}
\newblock \emph{\bibinfo{journal}{PASA}} \textbf{\bibinfo{volume}{13}}, \bibinfo{pages}{243--248}
  (\bibinfo{year}{1996}).

\bibitem{pfk15}
\bibinfo{author}{{Pietka}, M.}, \bibinfo{author}{{Fender}, R.~P.} \&
  \bibinfo{author}{{Keane}, E.~F.}
\newblock \bibinfo{title}{{The variability time-scales and brightness
  temperatures of radio flares from stars to supermassive black holes}}.
\newblock \emph{\bibinfo{journal}{Mon. Not. R. Astron. Soc.}} \textbf{\bibinfo{volume}{446}},
  \bibinfo{pages}{3687--3696} (\bibinfo{year}{2015}).

\bibitem{fbma15}
\bibinfo{author}{{Fong}, W.}, \bibinfo{author}{{Berger}, E..},
  \bibinfo{author}{{Margutti}, R.}, \& \bibinfo{author}{{Ashley}, B.~A.}
\newblock \bibinfo{title}{{A Decade of Short-duration Gamma-ray Burst Broad-band Afterglows: Energetics, Circumburst Densities, and Jet Opening Angles}}.
\newblock \emph{\bibinfo{journal}{Astrophys. J., submitted}} (\bibinfo{year}{2015}).
\newblock \eprint{1509.02922}.

\bibitem{berg14}
\bibinfo{author}{{Berger}, E.}
\newblock \bibinfo{title}{{Short-Duration Gamma-Ray Bursts}}.
\newblock \emph{\bibinfo{journal}{ARAA}} \textbf{\bibinfo{volume}{52}},
  \bibinfo{pages}{43--105} (\bibinfo{year}{2014}).

\bibitem{fkn+97}
\bibinfo{author}{{Frail}, D.~A.}, \bibinfo{author}{{Kulkarni}, S.~R.},
  \bibinfo{author}{{Nicastro}, L.}, \bibinfo{author}{{Feroci}, M.} \&
  \bibinfo{author}{{Taylor}, G.~B.}
\newblock \bibinfo{title}{{The radio afterglow from the {$\gamma$}-ray burst of
  8 May 1997}}.
\newblock \emph{\bibinfo{journal}{Nature}} \textbf{\bibinfo{volume}{389}},
  \bibinfo{pages}{261--263} (\bibinfo{year}{1997}).

\bibitem{shri15} \bibinfo{author}{{Kulkarni}, S.~R.},
  \bibinfo{author}{{Ofek}, E.~O.}, \& \bibinfo{author}{{Neill}, J.~D.}
  \newblock \bibinfo{title}{{The Arecibo Fast Radio Burst: Dense
      Circum-burst Medium}} (\bibinfo{year}{2015}).  \newblock
  \bibinfo{note}{Preprint (arXiv:astro-ph/1511.09137)}.

\setcounter{firstbib}{\value{enumiv}}
\end{thebibliography}

\newpage
\begin{thebibliography}{10}
\setcounter{enumiv}{\value{firstbib}}
\expandafter\ifx\csname url\endcsname\relax
  \def\url#1{\texttt{#1}}\fi
\expandafter\ifx\csname urlprefix\endcsname\relax\def\urlprefix{URL }\fi
\providecommand{\bibinfo}[2]{#2}
\providecommand{\eprint}[2][]{\url{#2}}

\bibitem{zok+14}
\bibinfo{author}{{Zheng}, Z.}, \bibinfo{author}{{Ofek}, E.~O.},
  \bibinfo{author}{{Kulkarni}, S.~R.}, \bibinfo{author}{{Neill}, J.~D.} \&
  \bibinfo{author}{{Juric}, M.}
\newblock \bibinfo{title}{{Probing the Intergalactic Medium with Fast Radio
  Bursts}}.
\newblock \emph{\bibinfo{journal}{Astrophys. J.}} \textbf{\bibinfo{volume}{797}},
  \bibinfo{pages}{71} (\bibinfo{year}{2014}).

\bibitem{den14}
\bibinfo{author}{{Dennison}, B.}
\newblock \bibinfo{title}{{Fast radio bursts: constraints on the dispersing medium}}
\newblock \emph{\bibinfo{journal}{Mon. Not. R. Astron. Soc.}} \textbf{\bibinfo{volume}{443}},
  \bibinfo{pages}{L11-L14} (\bibinfo{year}{2014}).

\bibitem{hsm04}
\bibinfo{author}{{Hotan}, A.~W.}, \bibinfo{author}{{van Straten}, W.} \&
  \bibinfo{author}{{Manchester}, R.~N.}
\newblock \bibinfo{title}{{PSRCHIVE and PSRFITS: An Open Approach to Radio
  Pulsar Data Storage and Analysis}}.
\newblock \emph{\bibinfo{journal}{PASA}} \textbf{\bibinfo{volume}{21}},
  \bibinfo{pages}{302--309} (\bibinfo{year}{2004}).

\bibitem{mwa13}
\bibinfo{author}{{Tingay}, S.~J.} \emph{et~al.}
\newblock \bibinfo{title}{{The Murchison Widefield Array: The Square Kilometre
  Array Precursor at Low Radio Frequencies}}.
\newblock \emph{\bibinfo{journal}{PASA}} \textbf{\bibinfo{volume}{30}},
  \bibinfo{pages}{7} (\bibinfo{year}{2013}).

\bibitem{ligo09}
\bibinfo{author}{{Abbott}, B.~P.} \emph{et~al.}
\newblock \bibinfo{title}{{LIGO: the Laser Interferometer Gravitational-Wave
  Observatory}}.
\newblock \emph{\bibinfo{journal}{Reports on Progress in Physics}}
  \textbf{\bibinfo{volume}{72}}, \bibinfo{pages}{076901}
  (\bibinfo{year}{2009}).

\bibitem{bhwp10}
\bibinfo{author}{{Becker}, R.~H.}, \bibinfo{author}{{Helfand}, D.~J.},
  \bibinfo{author}{{White}, R.~L.} \& \bibinfo{author}{{Proctor}, D.~D.}
\newblock \bibinfo{title}{{Variable Radio Sources in the Galactic Plane}}.
\newblock \emph{\bibinfo{journal}{AJ}} \textbf{\bibinfo{volume}{140}},
  \bibinfo{pages}{157--166} (\bibinfo{year}{2010}).

\bibitem{ofb+11}
\bibinfo{author}{{Ofek}, E.~O.} \emph{et~al.}
\newblock \bibinfo{title}{{A Very Large Array Search for 5 GHz Radio Transients
  and Variables at Low Galactic Latitudes}}.
\newblock \emph{\bibinfo{journal}{Astrophys. J.}} \textbf{\bibinfo{volume}{740}},
  \bibinfo{pages}{65} (\bibinfo{year}{2011}).

\bibitem{fko+12}
\bibinfo{author}{{Frail}, D.~A.}, \bibinfo{author}{{Kulkarni}, S.~R.},
  \bibinfo{author}{{Ofek}, E.~O.}, \bibinfo{author}{{Bower}, G.~C.} \&
  \bibinfo{author}{{Nakar}, E.}
\newblock \bibinfo{title}{{A Revised View of the Transient Radio Sky}}.
\newblock \emph{\bibinfo{journal}{Astrophys. J.}} \textbf{\bibinfo{volume}{747}},
  \bibinfo{pages}{70} (\bibinfo{year}{2012}).

\bibitem{cbw13}
\bibinfo{author}{{Croft}, S.}, \bibinfo{author}{{Bower}, G.~C.} \&
  \bibinfo{author}{{Whysong}, D.}
\newblock \bibinfo{title}{{The Allen Telescope Array Pi GHz Sky Survey. III.
  The ELAIS-N1, Coma, and Lockman Hole Fields}}.
\newblock \emph{\bibinfo{journal}{Astrophys. J.}} \textbf{\bibinfo{volume}{762}},
  \bibinfo{pages}{93} (\bibinfo{year}{2013}).

\bibitem{gop+06}
\bibinfo{author}{{Gal-Yam}, A.} \emph{et~al.}
\newblock \bibinfo{title}{{Radio and Optical Follow-up Observations of a
  Uniform Radio Transient Search: Implications for Gamma-Ray Bursts and
  Supernovae}}.
\newblock \emph{\bibinfo{journal}{Astrophys. J.}} \textbf{\bibinfo{volume}{639}},
  \bibinfo{pages}{331--339} (\bibinfo{year}{2006}).

\bibitem{mks+02}
\bibinfo{author}{{Miyazaki}, S.} \emph{et~al.}
\newblock \bibinfo{title}{{Subaru Prime Focus Camera -- Suprime-Cam}}.
\newblock \emph{\bibinfo{journal}{PASJ}} \textbf{\bibinfo{volume}{54}},
  \bibinfo{pages}{833--853} (\bibinfo{year}{2002}).

\bibitem{dib+14}
\bibinfo{author}{{Davenport}, J.~R.~A.} \emph{et~al.}
\newblock \bibinfo{title}{{The SDSS-2MASS-WISE 10-dimensional stellar colour
  locus}}.
\newblock \emph{\bibinfo{journal}{Mon. Not. R. Astron. Soc.}} \textbf{\bibinfo{volume}{440}},
  \bibinfo{pages}{3430--3438} (\bibinfo{year}{2014}).

\bibitem{cwm03}
\bibinfo{author}{{Cohen}, M.}, \bibinfo{author}{{Wheaton}, W.~A.} \&
  \bibinfo{author}{{Megeath}, S.~T.}
\newblock \bibinfo{title}{{Spectral Irradiance Calibration in the Infrared.
  XIV. The Absolute Calibration of 2MASS}}.
\newblock \emph{\bibinfo{journal}{AJ}} \textbf{\bibinfo{volume}{126}},
  \bibinfo{pages}{1090--1096} (\bibinfo{year}{2003}).

\bibitem{jcm+11}
\bibinfo{author}{{Jarrett}, T.~H.} \emph{et~al.}
\newblock \bibinfo{title}{{The Spitzer-WISE Survey of the Ecliptic Poles}}.
\newblock \emph{\bibinfo{journal}{Astrophys. J.}} \textbf{\bibinfo{volume}{735}},
  \bibinfo{pages}{112} (\bibinfo{year}{2011}).

\bibitem{bdc08}
\bibinfo{author}{{Brammer}, G.~B.}, \bibinfo{author}{{van Dokkum}, P.~G.} \&
  \bibinfo{author}{{Coppi}, P.}
\newblock \bibinfo{title}{{EAZY: A Fast, Public Photometric Redshift Code}}.
\newblock \emph{\bibinfo{journal}{Astrophys. J.}} \textbf{\bibinfo{volume}{686}},
  \bibinfo{pages}{1503--1513} (\bibinfo{year}{2008}).

\bibitem{cce08}
\bibinfo{author}{{da Cunha}, E.}, \bibinfo{author}{{Charlot}, S.} \&
  \bibinfo{author}{{Elbaz}, D.}
\newblock \bibinfo{title}{{A simple model to interpret the ultraviolet, optical
  and infrared emission from galaxies}}.
\newblock \emph{\bibinfo{journal}{Mon. Not. R. Astron. Soc.}} \textbf{\bibinfo{volume}{388}},
  \bibinfo{pages}{1595--1617} (\bibinfo{year}{2008}).

\bibitem{fj76}
\bibinfo{author}{{Faber}, S.~M.} \& \bibinfo{author}{{Jackson}, R.~E.}
\newblock \bibinfo{title}{{Velocity dispersions and mass-to-light ratios for
  elliptical galaxies}}.
\newblock \emph{\bibinfo{journal}{Astrophys. J.}} \textbf{\bibinfo{volume}{204}},
  \bibinfo{pages}{668--683} (\bibinfo{year}{1976}).

\bibitem{bks+01}
\bibinfo{author}{{Bullock}, J.~S.} \emph{et~al.}
\newblock \bibinfo{title}{{Profiles of dark haloes: evolution, scatter and
  environment}}.
\newblock \emph{\bibinfo{journal}{Mon. Not. R. Astron. Soc.}} \textbf{\bibinfo{volume}{321}},
  \bibinfo{pages}{559--575} (\bibinfo{year}{2001}).

\bibitem{ken98}
\bibinfo{author}{{Kennicutt}, R.~C., Jr.}
\newblock \bibinfo{title}{{The Global Schmidt Law in Star-forming Galaxies}}.
\newblock \emph{\bibinfo{journal}{Astrophys. J.}} \textbf{\bibinfo{volume}{498}},
  \bibinfo{pages}{541--552} (\bibinfo{year}{1998}).

\end{thebibliography}
\end{document}